\documentclass[12pt,preprint]{aastex}

\newcommand{\un}[1]{~\hspace{-1pt}\ensuremath{\mathrm{#1}}}
\newcommand{\titane}[1]{$^{44}{}${#1}}

\newcommand{\integ}{{\it INTEGRAL}~}
\newcommand{\ibis}{IBIS~}
\newcommand{\spi}{SPI~}
\newcommand{\isgri}{ISGRI~}

\newcommand{\chandra}{{\it Chandra}~}
\newcommand{\asca}{{\it ASCA}~}
\newcommand{\gro}{{\it GRO}}
\newcommand{\sax}{{\it BeppoSAX}}
\newcommand{\simbolx}{{\it SIMBOL-X}~}

\newcommand{\gammaray}{$\gamma$-ray~}

\newcommand{\xray}{X-ray~}

\def\d{$^\circ$}
\def\m{$^\prime$}
\def\s{$^{\prime\prime}$}

\def\cm3{cm$^{-3}$}

\def\eg{{\it e.g.~}}
\def\etal{et~al.~}

\begin{document}

\title{G0.570-0.018: A young supernova remnant ?\\
       {\bf INTEGRAL} and VLA observations}

\slugcomment{Revised version for ApJ on 07 October 2005}
\shorttitle{\gammaray and radio observations of G0.570-0.018}
\shortauthors{Renaud \etal}

\author{
    M. Renaud\altaffilmark{1,2}, S. Paron\altaffilmark{3}, R. Terrier\altaffilmark{2,1},
    F. Lebrun\altaffilmark{1,2}, G. Dubner\altaffilmark{3}, E. Giacani\altaffilmark{3},
    A.M. Bykov\altaffilmark{4}
    }

\altaffiltext{1}{\scriptsize Service d'Astrophysique,
DAPNIA/DSM/CEA, 91191 Gif-sur-Yvette, France; mrenaud@cea.fr}

\altaffiltext{2}{\scriptsize APC-UMR 7164, 11 place M. Berthelot,
75231 Paris, France}

\altaffiltext{3}{\scriptsize Instituto de Astronomia y Fisica del
Espacio (IAFE), CC 67, Suc. 28, 1428 Buenos Aires, Argentina}

\altaffiltext{4}{\scriptsize A.F. Ioffe Institute for Physics and 
Technology, St. Petersburg, Russia, 194021}

\begin{abstract}

We report {\it INTEGRAL}/\ibis \gammaray and VLA radio observations of G0.570-0.018, a diffuse \xray source recently discovered by 
\asca and \chandra in the Galactic center region. Based on its spectrum and morphology, G0.570-0.018 has been proposed to be a very 
young supernova remnant. In this scenario, the presence of  \gammaray lines coming from the short-lived radioactive nucleus \titane{Ti} 
as well as synchrotron radio continuum emission are expected. The first could provide informations on nucleosynthesis environments
in the interior of exploding stars, the latter could probe the interaction between the supernova blast wave and the 
circumstellar/interstellar matter. We have not detected \titane{Ti} lines nor any conspicuous radio feature associated with this 
source down to the achieved sensitivities. From the derived upper limits we set constraints on the nature of G0.570-0.018. 

\end{abstract}



\section{Introduction}
\label{s:intro}

The Galactic center region is one of the richest regions in sources of the Milky Way. Numerous \xray binaries, massive stellar
clusters and supernova remnants (hereafter, SNRs) lie in this complex region, immersed in an extended high-temperature plasma which 
heavily dominates the global soft \xray emission \citep{c:koyama96}. Recently, a diffuse \xray source, G0.570-0.018, was detected 
with \asca and confirmed with \chandra observations. The X-ray source has a ring-like structure, about 10\s~in radius, with a 
spectrum that can be fitted by a thermal emission model with a temperature of about 6\un{keV} \citep{c:senda02}. The high value 
derived for N$_{H}$ ($\sim$ 10$^{23}$ cm$^{-2}$) is consistent with a source located near the Galactic center, at a distance of about 
8 kpc. Based on the detection of an iron line with a high equivalent width at 6.5\un{keV} and a high Fe abundance (suggesting a recently
shocked plasma), the authors concluded that the origin must be a very young SNR with an age of $\sim$ 80 yr. However, based on 
the analysis in the \chandra image of a faint east-west X-ray tail possibly related to G0.570-0.018, they conclude that it 
could be about twice older. In a SNR scenario, we might expect non-thermal radio continuum emission from shock-accelerated electrons 
and the presence of \gammaray lines associated with the decay of \titane{Ti}. This radioactive nucleus is thought to be exclusively 
produced in supernova (hereafter, SN) explosions. It is primarily generated in the $\alpha$-rich freeze-out from nuclear statistical 
equilibrium occurring in the explosive silicon burning stage of core-collapse SNe (\eg Woosley \& Weaver 1995), while a normal 
freeze-out Si burning is at play in Type Ia SNe \citep{c:thielemann86}. The radioactive decay chain 
\titane{Ti}$\longrightarrow$\titane{Sc}$\longrightarrow$\titane{Ca}, with a lifetime of about 87.5 yr \citep{c:wietfeldt99}, 
produces three \gammaray lines at 67.9\un{keV}, 78.4\un{keV} (from \titane{Sc}$^{\star}$) and at 1157\un{keV} 
(from \titane{Ca}$^{\star}$) with similar branching ratios. It has been shown that \titane{Ti} may reveal young galactic SNRs 
that could be hidden from our view at optical wavelengths due to heavy extinction in the Galactic plane region. \titane{Ti} \gammaray 
lines have been observed in the young SNRs Cassiopeia~A (Iyudin \etal 1994; Vink \etal 2001; Vink 2005) and RXJ~0856-4622 
\citep{c:iyudin98}. Moreover, the radio data can provide information on the interaction of the blast wave with the circumstellar 
and local interstellar media. G0.570-0.018 is then an ideal case to search for specific signatures of young SNe such as the 
\titane{Ti} lines and the radio non-thermal radiation by combining both observations as early proposed by Silberberg \etal (1993).

In this paper, based on the \integ (IBIS/ISGRI, section \ref{s:isgri}) and VLA\footnote[1]{The Very Large Array of the National Radio
Astronomy Observatory is a facility of the NSF operated under cooperative agreement by Associated Universities Inc.} (section 
\ref{s:radio}) data, we investigate if G0.570-0.018 is indeed a young SNR. If this is the case, its possible type and characteristics 
are discussed in section \ref{s:discuss} at the light of all available observationnal and theoretical material.


\section{{\bf INTEGRAL} \& VLA observations}
\label{s:obs}

The INTErnational Gamma-Ray Astrophysics Laboratory \integ \citep{c:winkler03} is an ESA mission carrying two main telescopes 
based on a coded aperture imaging system: \ibis \citep{c:ubertini03} and \spi \citep{c:vedrenne03}. \ibis, with its soft \gammaray 
(15\un{keV} - 1\un{MeV}) imager \isgri \citep{c:lebrun03}, provides the finest imaging (13\m ~FWHM) and the best line sensitivity up 
to a few hundred\un{keV}. It is therefore best suited for a search for the two low energy \titane{Ti} lines and we only analyzed data 
from this instrument. \integ is mainly devoted to the observation of galactic sources and spends considerable amount of time pointing 
towards the Galactic plane and the Galactic center regions. We performed the analysis of data from the Galactic Center Deep Exposure 
(GCDE) of the Core Program, acquired during the first two years of \integ operations. In a search for the radio counterpart for 
G0.570-0.018, we have reprocessed archival VLA data at $\lambda$ 20\un{cm} and 6\un{cm} from different pointings where this source 
was within the field of view. We have also analyzed the 90\un{cm} intermediate resolution VLA image of the Galactic center from 
LaRosa \etal (2000).


\section{Analysis and Results}
\label{s:res}

\subsection{$^{44}$Ti \gammaray lines}
\label{s:isgri}
\clearpage
\begin{figure*}[htb]
\epsscale{1.0} 
\begin{center}
\includegraphics[scale=0.5, angle=0]{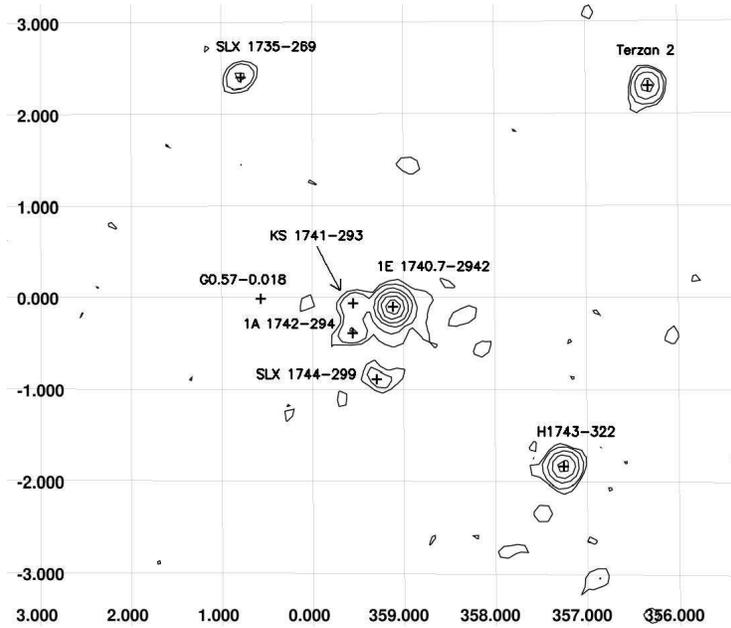}
\end{center}

\caption{IBIS/ISGRI contour map in Galactic coordinates of the combined significance
in the ranges of the two low energy \titane{Ti} lines. Contours are drawn from 3 to 100 
using a square root scale. The continuum emission of several known bright sources is 
clearly detected while there is no evidence of any emission coming from G0.570-0.018 
above the 3$\sigma$ level.\label{f:44Ti_ima}}

\end{figure*}
\clearpage
Both \asca and \chandra observations of G0.570-0.018 present a thermal \xray spectrum, without any evidence of synchrotron radiation. 
It is therefore highly improbable to detect this source with \ibis in the hard \xray continuum, above 15\un{keV}. Besides, as revealed 
by IBIS/ISGRI, the Galactic center region is a crowded area at hard \xray energies \citep{c:belanger05}, making it more difficult to 
precisely distinguish the part of the emission which could be associated with this source. In fact we did not find any evidence of 
such emission in the 20-40\un{keV} mosaic presented in B{\'e}langer \etal (2005). We analyzed pointings where G0.570-0.018 lies within 
the field of view of \ibis ($\le$ 15\d), in the 65-71 and 75-82\un{keV} energy bands centered on the two low energy \titane{Ti} 
\gammaray lines, using the Off-Line Scientific Analysis (OSA) software version 4.2 \citep{c:goldwurm03}. These narrow bandwidths take 
full advantage of the \isgri energy resolution ($\sim$ 6\un{keV} FWHM at 70\un{keV}). The critical point in the analysis of the 
IBIS/ISGRI data is the background subtraction: we have generated background shadowgrams (detector image containing the shadow of the 
coded mask onto the \isgri detector) by analyzing and summing a large part of the high-latitude and empty field observations performed 
during the first two years of the mission in 256 energy bands. The high total exposure time ($\sim$ 2\un{Ms}) warrants the best 
removal of structures in the detector images \citep{c:terrier03}, mainly around the K$_{\alpha}$ and K$_{\beta}$ fluorescence lines of 
the W (59 and 65\un{keV}) and Pb (75 and 85\un{keV}) located close to the two low energy \titane{Ti} astrophysical lines. This 
method provides flat background-subtracted detector images in any desired energy band. Thus, the convolution of these shadowgrams 
with a decoding array derived from the spatial characteristics of the coded mask produces good quality reconstructed sky images. We 
finally obtained two mosaic images in these two energy bands and combined both to increase the signal-to-noise ratio. The resulting 
mosaic is shown in Figure \ref{f:44Ti_ima}, with a final exposure time towards the source of about 4.3 Ms.

\begin{figure}[htb]
\epsscale{1.0} 
\begin{center}
\includegraphics[scale=0.45, angle=0]{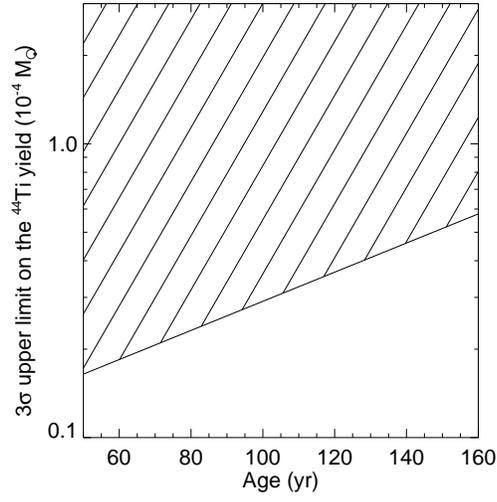}
\end{center}

\caption{3$\sigma$ upper limit on the \titane{Ti} yield (in units of 
10$^{-4}$~$\mathrm{M}_{\odot}$) as a function of the age (in years). 
The dashed region corresponds to the range of parameters excluded at 
the 3$\sigma$ confidence level.\label{f:y_age}}

\end{figure}

We found no evidence for any emission above 3$\sigma$ from G0.570-0.018 in the \titane{Ti} \gammaray lines range and estimated an 
upper limit of 1.2 $\times$ 10$^{-5}$ cm$^{-2}$ s$^{-1}$ at the 3$\sigma$ confidence level. This can be converted into an upper limit 
on the \titane{Ti} yield via the following equation:

\begin{equation}
Y_{44}=1.38\times{\left[\frac{F_{\gamma}}{\mathrm{cm}^{-2}~\mathrm{s}^{-1}}\times{\left(\frac{d}{1~\mathrm{kpc}}\right)^{2}}\times{e^{t/\tau_{44}}}\times{\frac{\tau_{44}}{1~\mathrm{yr}}}\right]\times{10^{-4}~\mathrm{M}_{\odot}}}
\end{equation}

where Y$_{44}$ is the \titane{Ti} yield, F$_{\gamma}$ the \titane{Ti} line flux and $\tau_{44}$ the \titane{Ti} lifetime $\sim$ 87.5 yr 
\citep{c:wietfeldt99}. Assuming that G0.570-0.018 is located near the Galactic center region, a distance of 8 kpc is adopted 
\citep{c:eisenhauer03}. Based on this assumption, we obtained the relation between the maximal \titane{Ti} yield and the age of 
this source presented in Figure \ref{f:y_age}. 
\clearpage
\clearpage
For an age of 80 yr, our 3$\sigma$ upper limit on the \titane{Ti} ejected mass is $\sim$ 2 $\times$ 10$^{-5}$ $\mathrm{M}_{\odot}$.
One should notice that this value is $\sim$ 8 times lower than that of the youngest known SNR Cassiopeia~A (hereafter, Cas~A) derived 
from \gro/COMPTEL \citep{c:iyudin94}, \sax/PDS \citep{c:vink01} and IBIS/ISGRI \citep{c:vink05} observations. This result is 
discussed in detail in section \ref{s:discuss}.

\subsection{Radio emission}
\label{s:radio}
\clearpage
\begin{figure*}[htb] 
\epsscale{1.0}  
\begin{center} 
\includegraphics[scale=0.8, angle=0]{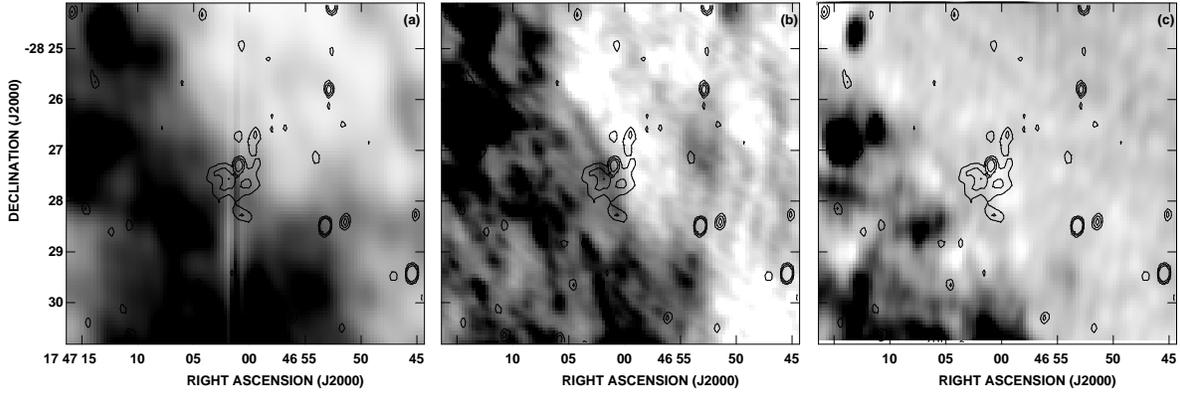} 
\end{center} 
 
\caption{Radio images of a $\sim$ 7\m~field around G0.570-0.018 at 
$\lambda$ 90\un{cm} {\bf (a)}, 20\un{cm} {\bf (b)} and 6\un{cm} {\bf (c)}.  
The contours represent X-ray emission convolved with a gaussian beam  
of 13\s $\times$ 8\s. The greyscale varies between 0 and 150 mJy/beam  
at 90\un{cm}, between -5 and 10 mJy/beam at 20\un{cm}, and between -10 
and 40 mJy/beam at 6\un{cm}.\label{f:radio_im}} 

\end{figure*}
\clearpage
The image at 20\un{cm} was obtained from archival data corresponding to observations carried out using the VLA in the hybrid CnB 
configuration for about 1.5 hours on 2004 February. The data were processed under the MIRIAD software package \citep{c:sault95} 
following standard procedures. The resulting synthesized beam is 13\s $\times$ 8\s, PA=54.6, and the average rms noise of the order 
of 2.5 mJy/beam. The largest scale structure accessible to this array is of about 5\m. A field around G0.570-0.018 is shown in Figure 
\ref{f:radio_im}b, with a few contours superimposed representing the \chandra \xray emission smoothed to the same angular resolution 
as the radio image. Correction for primary beam attenuation was applied.

To produce the image at 6\un{cm} we used archival data from observations carried out with the VLA-D array during 45 minutes on 2003 
April 30. The synthesized beam is 27\s $\times$ 11\s, PA=-5.2 and the average rms noise in the field, about 2.2 mJy/beam. The 
resulting image is displayed in Figure \ref{f:radio_im}c, again with \chandra \xray contours overlapped.

The image shown in Figure \ref{f:radio_im}a at 90\un{cm} was extracted from the 4\d $\times$ 4\d~image of the center of the Galaxy 
obtained with the VLA B, C and D configurations (LaRosa et al. 2000). The angular resolution of these data is 48\s $\times$ 
48\s~and the rms sensitivity of 5.9 mJy/beam. The largest angular scale to which this image is sensitive is approximately 45\m.

From these images, it is apparent that G0.570-0.018 lies on a region with tenuous, smooth radio emission. Not any conspicuous 
radio feature can be associated with the \xray source at any frequency up to the limit of the angular resolution and sensitivity 
of these images. An upper limit for the flux densities at 90 and 20\un{cm} can be estimated by integrating the radio emission over a 
region with the size of the outer \xray contour. From this integration we estimate: S$_{\rm 90\un{cm}} \sim 0.23$ Jy and 
S$_{\rm 20\un{cm}} \sim 0.066$ Jy. Although the flux contribution from large-scale structures was not added to these images, the flux 
density estimates are reliable within observational errors because the size of the studied structure is smaller than the largest well 
imaged structure at the respective frequencies. This is not the case for the emission at 6\un{cm}, thus it was not possible to 
accurately estimate the flux density at this short wavelength.


\section{Discussion}
\label{s:discuss}

The \xray morphology and spectra of G0.570-0.018 show a hot-temperature plasma distributed in a ring-like structure, two 
characteristic features of young ejecta-dominated SNRs. In this context, we attempted to shed further light on the nature of this 
source searching for  \titane{Ti} \gammaray lines and radio continuum emission. However, to our surprise neither the \titane{Ti} 
\gammaray lines observations nor the radio continuum maps reveal any feature that can be ascribed to G0.570-0.018 at the levels of 
sensitivity presented above. 

\begin{figure}[htb]
\epsscale{1.0} 
\begin{center}
\includegraphics[scale=0.5, angle=0]{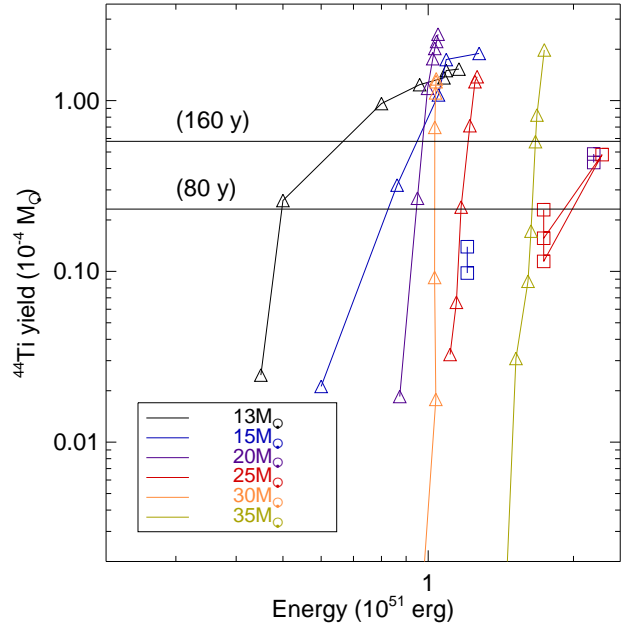}
\end{center}

\caption{\titane{Ti} yield (in units of 10$^{-4}$~$\mathrm{M}_{\odot}$)
versus the energy of the explosion (in units of 10$^{51}$ erg) for two 
sets of nucleosynthesis calculations in Type II SNe: triangles represent 
those of Limongi \& Chieffi (2003) and squares those of Rauscher \etal (2002). 
Each color is related to the mass of the progenitor. Horizontal lines
represent our 3$\sigma$ upper limit for two different ages of G0.570-0.018.
\label{f:models}}

\end{figure}

Concerning SNe nucleosynthesis products such as \titane{Ti}, presupernova evolutions and calculations of explosive yields have a long
history. Recently, core-collapse SNe have been studied by Woosley \& Weaver (1995), Thielemann \etal (1996), Rauscher \etal (2002, 
hereafter RHHW02) and Limongi \& Chieffi (2003, hereafter LC03) while nucleosynthesis in Chandrasekhar mass models for Type Ia SNe can 
be found in Iwamoto \etal (1999). The \titane{Ti} yields used in this paper were extracted from those obtained by RHHW02 and LC03 
since their calculations include several improvements in stellar physics and revised nuclear reaction rates. Figure \ref{f:models} 
shows these \titane{Ti} yields as a function of the energy of the explosion, for different masses of the progenitor with our 
3$\sigma$ upper limit for the two possible ages of G0.570-0.018.
\clearpage
\clearpage
Explosive yields are sensitive to many details of the explosion and in the case of core-collapse SNe, \titane{Ti} is thought to 
be created close to the mass-cut (the mass above which the matter falls back onto the compact remnant). From Figure \ref{f:models} it 
can be seen that for each track of constant progenitor mass, the higher is the energy of the explosion, the larger is the ejected 
mass of \titane{Ti}. For an age of 80 yr, our 3$\sigma$ upper limit is only compatible with few models of core-collapse SNe, where 
the energy is unsufficient to eject an amount of \titane{Ti} above 2 $\times$ 10$^{-5}$~$\mathrm{M}_{\odot}$. We also considered the 
expected \titane{Ti} yields in various Chandrasekhar mass models for SNIa explosions (Iwamoto \etal 1999 and references therein). In 
these models, explosive nucleosynthesis is calculated for a variety of deflagration speeds and ignition densities. We find that 
``standards'' models (Nomoto \etal 1984; Thielemann \etal 1986) predict \titane{Ti} yields clearly below our 3$\sigma$ upper limit, 
while delayed-detonation models with highly energetic explosions ($>$ 1.4 $\times$ 10$^{51}$ ergs) predict \titane{Ti} yields between 
$\sim$ 3 and 4.5 $\times$ 10$^{-5}$~$\mathrm{M}_{\odot}$ which could be compatible only in the case where G0.570-0.018 is at least 
two times older ($>$ 160 yr). The situation is worse in the case of sub-Chandrasekhar SNIa explosions where helium burns in a 
detonation wave causing a detonation also in the interior of the white dwarf (Woosley \etal 1986; Woosley \& Weaver 1994). In such 
systems, substantial overproductions of \titane{Ti} from several 10$^{-4}$~$\mathrm{M}_{\odot}$ up to 4 $\times$ 
10$^{-3}$~$\mathrm{M}_{\odot}$ are expected and then well above our sensitivity. In summary, comparing our upper limits with various 
models, we conclude that only few sub-energetic core-collapse or standard thermonuclear explosions predict \titane{Ti} yields 
that explain our non-detection at the IBIS/ISGRI sensitivity. 

From an observational point of view, the 1.157\un{MeV} \titane{Ti} \gammaray line was detected for the first time in Cas~A with 
\gro/COMPTEL \citep{c:iyudin94}. Later, Vink \etal (2001) reported the detection with \sax/PDS of the two low energy \titane{Ti} 
lines in this SNR implying an initial \titane{Ti} mass of (0.8-2.5) $\times$ 10$^{-4}$~$\mathrm{M}_{\odot}$. Preliminary analysis 
of the IBIS/ISGRI data on Cas~A yielded a detection of the 68 keV \titane{Ti} line with a flux consistent with the \sax~detections
\citep{c:vink05}. Moreover, it is suspected that the late-time light curve of SN~1987A ($\succeq$ 2000 days) is dominated by the 
\titane{Ti} decay. From time-dependent models for the light curve in combination with broad band photometry, Fransson \& Kozma (2001) 
estimated a \titane{Ti} mass of (0.5-2.0) $\times$ 10$^{-4}$~$\mathrm{M}_{\odot}$. On the other hand, Senda \etal (2002) noticed 
similarities between the \xray structure of the ring of G0.570-0.018 and that of SN~1987A where strong stellar winds might produce a 
gas ring being heated by SN ejecta. However, one can see from Figure \ref{f:y_age} that G0.570-0.018 should have been detected in the 
\titane{Ti} \gammaray lines range if it had emitted yields comparable to those measured from Cas~A or those inferred from SN~1987A, 
even in the worst case of considering the lower limits of \titane{Ti} yields, and the oldest age of 160 yr.  

Concerning the radio observations, one can expect synchrotron emission coming from either initially accelerated or shock-accelerated 
electrons and the non-observation at any radio frequency of G0.570-0.018 is then surprising. If G0.570-0.018 has an age of 80 yr, it 
should be considered as an ``intermediate-age'' supernova (see Eck \etal 2002 and references therein) evolving from SN to SNR. 
G0.570-0.018, however, could be as old as 160 yr. With such an upper limit on the age, we cannot exclude that G0.570-0.018 is in the 
SNR phase where the electrons responsible for the radio synchroton emission are accelerated at the shock front. The main difficulty 
of the studies of radio emission from SNe/SNRs is that between the observed extragalactic radio SNe (hereafter, RSNe) and the youngest 
known Galactic SNR (Cas~A), there is an observationnal gap of about 300 yr and then the evolution from RSN to SNR is poorly 
understood. Type II SNe seem to be well described by Chevalier's models (Chevalier 1982a, 1982b, 1984) where the radio emission 
is related to the circumstellar matter while that of older SNe, entering the SNR phase (predicted to take at least 100 yr), can be 
related to the amount of interstellar matter (Gull 1973; Cowsik \& Sarkar 1984). Emission mechanisms in both cases are anyway 
identical. Thus, we have divided our discussion in two parts, one considering that G0.570-0.018 is still a SN and an other assuming 
that it could be a very young SNR.

In the first case, according to the Chevalier's model \citep{c:chevalier82b}, the flux of RSNe drastically depends on the ratio 
between the mass-loss rate of the pre-SN progenitor and the wind speed. For small values, no strong radio emission is expected, thus 
explaining why despite several searches (\eg Eck \etal 1995), no Type Ia SNe have been detected in radio and why most of the 
extragalactic SNe seen at radio wavelenghts (Weiler \etal 2002) have progenitors with dense winds. Massive stars are thought to 
experience strong mass losses during the final stages of their evolution implying that they would be bright radio sources. Our 
non-detection in radio therefore points to the lower progenitor masses. Figure \ref{f:models} shows that for these lower masses, our 
upper limit on the production of \titane{Ti} favours rather low energetic events. On the other hand, G0.57-0.018 could be similar to 
the Type Ia SN~1885A in M31, the first extragalactic SN identified, where no radio emission has been found \citep{c:crane92}.

Now if we assume that G0.57-0.018 has an age of 160 yr, it can be considered as a very young SNR \citep{c:cowsik84}. 
Recently, Asvarov (2000) developed a model based on the diffusive shock acceleration mechanism to explain radio emission of adiabatic 
SNRs and compared its modeled Surface Brightness - Diameter ($\Sigma$ - D) evolutionary tracks with the empirical $\Sigma$ - D diagram.
He found that adiabatic SNRs evolve at nearly constant surface brightness: $\Sigma$ $\propto$ D$^{-0.5}$. We have used this relation 
and compared the surface brightness of the historical SNRs Kepler, Tycho, Cas~A and SN~1006 (we exclude the Crab nebula and 3C58 
because they are continuously fed by relativistic electrons from a central neutron star) as given by Green (2005) and scaled to an age 
of 160 yr, with our upper limit at 20\un{cm}, $\Sigma \sim$ 2 $\times$ 10$^{-20}$ W m$^{-2}$ Hz$^{-1}$ sr$^{-1}$. The only concordance 
found is with SN1006, suggesting that G0.570-0.018 might be a Type Ia SNR evolving in a low-density medium. This is consistent with 
our upper limit on the \titane{Ti} yield, since standard Type Ia SNe produce in average less \titane{Ti} than core-collapse SNe, and 
with its very low radio flux density. Although SNIa can be found at any location in spiral galaxies like the Milky Way, the location of 
G0.570-0.018 within the disk and close to the Galactic center, which is known to host abundant molecular material and young stars 
\citep{c:figer04}, makes the scenario of a core-collapse event more likely. In any case G0.570-0.018 must be located in a region 
probably depleted by a succession of strong stellar winds of massive stars and/or previous SN explosions. 

In any scenario of a SN/SNR, a possible explanation for the lack of radio emission can be the fact that G0.570-0.018 appears to be 
evolving in a low density hot plasma. In such medium high Mach numbers shocks may not be formed, even if velocities are high, as the 
sound speed in a hot plasma is also high. Such SNR shocks will not give rise to high compression ratios, and also shock acceleration 
may be less efficient. Therefore, the radio emission may not be very strong. In addition, if the SNR shock expands within a bubble 
blown out by the wind of the precursor star, the possibility of creating a radio synchrotron shell is even lower.

\section{Conclusion}
\label{s:conclusion}

In conclusion, the present \gammaray and radio observations have not solved the basic question if G0.57-0.018 is a genuine SN/SNR. If
it is, then our upper limits on the ejected \titane{Ti} mass and radio emission from G0.57-0.018 help to constrain its 
characteristics: the weak production of \titane{Ti} rules out all the sub-Chandrasekhar Type Ia SN scenarios and can be explained if 
G0.57-0.018 was a sub-energetic core-collapse supernova from a moderate mass progenitor or a standard thermonuclear explosion. Since 
no Type Ia SNe in radio have been detected, the second scenario looks more likely although the location of G0.57-0.018, very close to 
the Galactic center, might suggest a massive star progenitor. According to Chevalier's model, the very weak radio surface 
brightness is probably due to the low-density surrounding medium. If the SN/SNR nature of G0.57-0.018 is questioned, then the \xray 
morphology, the high Fe abundance and the position and width of the Fe line observed by Senda \etal (2002) remain to be explained. 
In any scenario of a young SN/SNR, G0.57-0.018 seems to be unusual and only the next generation of hard \xray/soft \gammaray focusing 
telescopes, such as \simbolx \citep{c:ferrando04}, will be in position to disentangle its nature.


\acknowledgements{M.R. gratefully thanks J. Paul for fruitful discussions and his various suggestions. The present work was supported 
		 with action ECOS A04U03 and based on observations with \integ, an ESA project with instruments and science data 
		 center (ISDC) funded by ESA members states (especially the PI countries: Denmark, France, Germany, Italy, Switzerland, 
		 Spain, Czech Republic and Poland, and with the participation of Russia and the USA). \isgri has been realized and 
		 maintained in flight by CEA-Saclay/DAPNIA with the support of CNES. S. P. is fellow of CONICET (Argentina). G. D. and 
		 E. G. are members of the {\it Carrera del Investigador Cient\'\i fico} of CONICET (Argentina). This research was 
		 partially funded by the UBACYT Grant A055 and by ANPCyT-PICT04-14018 (Argentina). A.M.B. was partially supported by 
		 RBRF 03-04-17433 and 04-02-16595.}

\end{document}